\documentclass[aps,prd,twocolumn,nofootinbib,floatfix,superscriptaddress,showpacs]{revtex4-1}
\usepackage[latin1]{inputenc}  
\usepackage{bbm,slashed}
\usepackage{mathrsfs}
\usepackage{graphicx,epsfig}
\graphicspath{{figures/}}
\usepackage{amsmath,amsfonts,amssymb}
\usepackage{booktabs} 
\usepackage{dsfont}
\usepackage{bm,bbm}

 \usepackage{enumitem}

\newcommand{\N}{\mathcal{N}} 
\newcommand{\vect}[1]{{\bm{#1}}}
\newcommand{\mbar}{{m}}
\newcommand{\gbar}{{g}}
\newcommand{\zbar}{\bar{z}} 

\newcommand{\Fbar}{\bar{F}}
\newcommand{\parbar}{\bar{\partial}} 
\newcommand{\phibar}{\bar{\phi}}
\newcommand{\psibar}{\bar{\psi}}
\newcommand{\Wbar}{\bar{W}}
\newcommand{\ubar}{\bar u}
\newcommand{\mcor}{m_{\rm corr}}

\newcommand{\abs}[1]{\left| #1 \right|}

\DeclareMathOperator{\STr}{STr} 

\DeclareMathOperator{\Str}{Str}

 \DeclareMathAlphabet{\boldmathe}{T1}{cmr}{bx}{it}

\begin{document}
\title{The two dimensional $\N=(2,2)$ Wess-Zumino Model in the Functional
Renormalization Group Approach}
\author{Franziska Synatschke-Czerwonka}
\affiliation{
Theoretisch-Physikalisches Institut, Friedrich-Schiller-Universit{\"a}t
Jena,
Max-Wien-Platz~1, D-07743~Jena, Germany} 
\author{Thomas Fischbacher}
\affiliation{University of Southampton, School of Engineering
Sciences, Highfield Campus, University Road, SO17 1BJ Southampton,
United Kingdom}
\author{Georg Bergner}
\affiliation{Institut~f{\"u}r~Theoretische Physik, Westf{\"a}lische~Wilhelms-Universit{\"a}t~M{\"u}nster, Wilhelm-Klemm-Str.~9, 48149~M{\"u}nster, Germany}

\begin{abstract}
	We study the supersymmetric $\N=(2,2)$ Wess-Zumino model in two dimensions with
	the functional renormalization group. At leading order in the
	supercovariant derivative expansion we recover the nonrenormalization theorem
	which states that the superpotential has no running couplings. 
	Beyond leading order the renormalization of the bare mass is
	caused by a momentum-dependent wave function renormalization.  To deal
	with the partial differential equations  we have developed a
	numerical toolbox called FlowPy. For weak couplings the quantum corrections to
	the bare mass found in lattice simulations are reproduced with high accuracy.
	But in the regime with intermediate couplings 	higher-order-operators that are
	not constrained by the nonrenormalization theorem  yield the dominating contribution to the 
	renormalized mass.
\end{abstract}
\maketitle

\section{Introduction}

In the search for high energy theories beyond the standard model
supersymmetric models are a topic of great interest. Supersymmetry reduces the
hierarchy and the fine-tuning problem. It has to be broken at
some energy scale since supersymmetry has not
been observed in low energy physics. The breaking does not occur on the
perturbative level  and therefore
nonperturbative tools are needed to analyze these models, e.\,g. lattice
formulations or the function renormalization group.

Lattice formulations and simulations have been
successfully applied to nonperturbative problems in field
theory. Although there has been considerable progress in the last years
\cite{Catterall:2009it,Giedt:2006pd,Bergner:2007pu,Kastner:2008zc} there are
still some difficulties in the lattice formulation of supersymmetry. The
discretization leads to a (partial) breaking of supersymmetry and the
implementation  of dynamical fermions on the lattice still poses a challenge.

Nonperturbative continuum methods, such as the functional
renormalization group (FRG) which manifestly preserve supersymmetry, can 
complement the lattice calculations. The FRG has previously been 
applied to a wide range of nonperturbative problems such as critical phenomena, fermionic systems,
gauge theories and quantum gravity, see  e.\,g.
\cite{Aoki:2000wm,Berges:2000ew,Litim:1998nf,Pawlowski:2005xe,Gies:2006wv,Sonoda:2007av,Rosten:2008ih,Rosten:2010vm,Delamotte:2003dw}
for reviews. Applied to supersymmetric theories it circumvents problems of the
lattice formulation such as supersymmetry breaking due to
discretization. But in order to solve the FRG equations truncations have to
be employed which introduce a different kind of error.

Quite a few conceptual studies of supersymmetric theories in the
framework of the  FRG have already been performed. The main ingredient
is the construction and use of a manifestly supersymmetric regularization
scheme. For example such a regulator has been presented for the four-dimensional Wess-Zumino model  in
\cite{Vian:1998kv,Bonini:1998ec}. Investigations for one-, two- and three-dimensional $\N=1$ Wess-Zumino models have been performed in
\cite{Synatschke:2008pv,Gies:2009az,Synatschke:2009nm,Synatschke:2010ub}.  A
FRG formulation of supersymmetric Yang-Mills theory  employing the
superfield formalism has been  given in \cite{Falkenberg:1998bg}; for further
applications see also \cite{Arnone:2004ey,Arnone:2004ek}.  General theories of 
a scalar superfield including the Wess-Zumino model  were studied with a
Polchinski-type RG equation in \cite{Rosten:2008ih},  which yields a new
approach to supersymmetric nonrenormalization theorems. The nonrenormalization
theorem has also been proven with FRG methods  in \cite{Sonoda:2009df}. In
\cite{Sonoda:2008dz} a Wilsonian effective action for  the Wess-Zumino model by
perturbatively iterating the FRG is constructed.

The aim of this work is twofold. On the one hand, we want to compare the
results from the supersymmetric formulation of the FRG equations to 
 lattice data  for the renormalized mass in the two-dimensional
$\N=(2,2)$ Wess-Zumino model \cite{Kastner:2008zc}. This comparison allows us
to estimate the truncation error. The renormalized mass is defined as the
location of the pole of the  propagator in the complex plane therefore we have
to take the momentum dependence in the FRG framework into account.

There are several applications, where this dependence is important but the related contributions lead to a much higher numerical effort 
for the solution of the flow equations.
Full momentum dependence of
propagators and vertices  has previously been
treated successfully in the literature
\cite{Ellwanger:1995qf,Pawlowski:2003hq,Fischer:2004uk,Blaizot:2006vr,Blaizot:2005wd,Blaizot:2005xy,Benitez:2009xg,Diehl:2007xz}.
We introduce a numerical toolbox called FlowPy which is designed
 to solve the flow equations with momentum dependence as encountered in this
 paper. FlowPy  can be adapted to solve not only flow
 equations with momentum dependence but also other differential equations
 encountered in the FRG framework e.\,g. for  field dependent effective
 potentials. In this paper we  demonstrate  that FlowPy solves
the flow equations reliably.

The paper is organized as follows:
In  section \ref{sec:Model} we introduce the $\N=(2,2)$ Wess-Zumino model in two
dimensions. In  section \ref{sec:flow} we sketch the derivation of the supersymmetric
flow equations for the superpotential and the (momentum-dependent) wave function
renormalization. The flow equation for the superpotential will lead to the
nonrenormalization theorem. In section \ref{sec:NumRes} first 
FlowPy is described and then we specialize our flow equations and demonstrate
that perturbation theory is reproduced correctly. In section \ref{sec:results}
we compare the renormalized mass calculated in the FRG approach with the results from lattice
simulations.

\section{The $\N=(2,2)$ Wess-Zumino model in two dimensions}
\label{sec:Model}
The $\N=(2,2)$ Wess-Zumino model in two dimensions can be found by a dimensional
reduction of the $\N=1$ model in four dimensions \cite{Wess:1973kz}. The Lagrange density  is given by
 \begin{align}
 \mathscr L=2\parbar\phibar\partial\phi+\psibar M\psi-\frac12\Fbar F
 		+\frac12FW'(\phi)+\frac12\Fbar\,\Wbar'(\phi)
 \end{align} 
with Dirac fermions $\psi$ and $\psibar$. The fermion matrix $M$ reads
\begin{align}
	M=\slashed{\partial}+W''(\phi)P_++\Wbar''(\phi)P_-
\end{align} 
with $P_{\pm}=(\mathds 1\pm\gamma_\ast)/2$, $F=F_1+iF_2$ and 
$\phi=\phi_1+i\phi_2$ as well as $\partial=\frac12(\partial_1-i\partial_2)$ and
$z=x_1+ix_2$. The superpotential is denoted by $W(\phi)=u(\phi_1,\phi_2)+iv(\phi_1,\phi_2)$.
 We work in the Weyl basis with $\gamma^1=\sigma_1,\gamma^2=-\sigma_2$ and
$\gamma_\ast=i\gamma^1\gamma^2=\sigma_3$. The complex spinors can be decomposed
as
$	\psi=\begin{pmatrix}
         	\psi_1& \psi_2
         \end{pmatrix}^{\rm T}$ and $\psibar=\begin{pmatrix}
            	\psibar_1&\psibar_2
            \end{pmatrix}.$
The Lagrange density is invariant under the supersymmetry transformations
\begin{equation}
  \begin{split} 
& \delta\phi=\psibar_1\varepsilon_1+\bar\varepsilon_1\psi_1,\;
 \delta\bar\phi=\psibar_2\varepsilon_2+\bar\varepsilon_2\psi_2,\\
& \delta\psibar_1=-\frac12F\bar\varepsilon_1-\partial\phi\bar\varepsilon_2,\;
 \delta\psibar_2=-\bar\partial\bar\phi\bar\varepsilon_1-\frac12\Fbar\bar\varepsilon_2,\\
& \delta\psi_1=-\frac12F\varepsilon_1+\bar\partial\phi\varepsilon_2,\;
 \delta\psi_2=\partial\bar\phi\varepsilon_1-\frac12\Fbar\varepsilon_2, \\
& \delta
F=2(\partial\psibar_1\varepsilon_2-\bar\varepsilon_2\bar\partial\psi_1),\;
\delta\Fbar=2(\partial\psibar_2\varepsilon_1-\bar\varepsilon_1\bar\partial\psi_2).
\end{split}
\label{eq:SuSyTrafos}
\end{equation}
The superspace formulation of this model is constructed in appendix 
\ref{sec:Superspace}. A detailed discussion of the underlying supersymmetry
algebra and a construction of the superspace
can be found e.\,g. in \cite{West}.

Integrating out the auxiliary fields yields
the on-shell Lagrangian
\begin{align}
\mathcal L_{\rm on}=2\bar\partial\phibar\partial\phi+\frac12 W'(\phi)\bar
W'(\phi) +\psibar M\psi.
\end{align} 
In this paper we will consider the superpotential 
\begin{align}W(\phi)=\frac12 m\phi^2+\frac13 g\phi^3.\label{eq:spot}\end{align}
The system has  two bosonic ground states which lead  to a nonzero Witten
index \cite{Witten:1982df}, therefore supersymmetry is never spontaneously
broken in the $\N=(2,2)$ Wess-Zumino model.

A characteristic feature of the $\N=1$ Wess-Zumino model in four dimensions
survives the dimensional reduction, namely 
that bosonic and fermionic loop corrections cancel in such a way that
the effective superpotential receives no quantum corrections. This is called
the nonrenormalization theorem
\cite{Witten:1993yc,MirrorSymmetry,Gates:1983nr}. In the two-dimensional model
considered here the cancellations even render the model finite. 
The model has been studied intensively in the literature, see e.\,g. 
\cite{Kastner:2008zc,Bergner:2007pu,Beccaria:1998vi,Catterall:2003ae,Giedt:2005ae} for 
 lattice simulations.

\section{Supersymmetric RG flow}
\label{sec:flow}
Following the lines of our previous works 
\cite{Synatschke:2010ub,Synatschke:2008pv,Synatschke:2009nm,Gies:2009az} we  construct a manifestly supersymmetric flow equation in
the
off-shell formulation.
Our approach is based on the FRG formulated in
terms of a flow equation for  the effective average action
$\Gamma_k$, i.e. the Wetterich equation  
\cite{Wetterich:1992yh}  
\begin{equation}
 \partial_k\Gamma_k=
 \frac12 \STr\left\{\left[\Gamma_k^{(2)}+ R_k\right]^{-1}\partial_k 
R_k\right\}.
\label{eq:lpa1}
\end{equation}
The scale dependent  $\Gamma_k$  interpolates
between the microscopic action $S$ for $k\to\Lambda$, with
$\Lambda$ denoting the microscopic UV scale, and the full quantum effective
action $\Gamma=\Gamma_{k\to0}$. As the  model considered in this paper is UV-finite, the cutoff $\Lambda$ can be set to infinity. The interpolating scale $k$
denotes an infrared (IR) regulator scale below which all fluctuations with
momenta smaller than $k$ are suppressed. For $k\to0$, all fluctuations are
taken into account  and we arrive at the full solution of the quantum theory in terms of the effective action $\Gamma$. The Wetterich equation defines an RG trajectory in the
space of action functionals with the classical action $S$ serving as initial
condition.

The second functional derivative of $\Gamma_k$ in
 Eq.~(\ref{eq:lpa1}) is defined as  
\begin{equation}
\left(\Gamma_k^{(2)}\right)_{ab}=\frac{\overrightarrow{\delta}}{\delta\Psi_a}
\Gamma_k\frac {\overleftarrow{\delta}}{\delta\Psi_b}\,,\label{eq:lpa3}
\end{equation}
where the indices $a,b$ summarize field components, internal and Lorentz
indices, as well as spacetime or momentum coordinates. In the present case, we
have $\Psi^{\text T}=(\phi,\phibar,F,\bar F,\psibar,\psi)$ where $\Psi$ is not
a superfield, but merely a collection of fields. The momentum-dependent regulator function
$R_k$ in Eq.~(\ref{eq:lpa1}) establishes the IR suppression of modes below
$k$. In the general case, three properties of the regulator $R_k(p)$ are
essential: (i) $R_k(p)|_{p^2/k^2\to 0} >0$ which implements the IR
regularization, (ii) $R_k(p)|_{k^2/p^2\to 0} =0$ which guarantees that the
regulator vanishes for $k\to0$, (iii) $R_k(p)|_{k\to\Lambda\to\infty}\to
\infty$ which serves to fix the theory at the classical action in the UV.
Different functional forms of $R_k$ correspond to different RG trajectories
manifesting the RG scheme dependence but the end point $\Gamma_{k\to 0}\to
\Gamma$ remains invariant, see e.~g.  
Refs.~\cite{Litim:2000ci,Litim:2001up,Litim:2001hk,Litim:2002cf,Pawlowski:2005xe,Litim:2006ag,Canet:2002gs}.
Supersymmetry is preserved if the regulator contribution to
the cutoff action $\Delta S_k$ (cf.~Eq.~\eqref{eq:CutoffAction}) is
supersymmetric.

As an ansatz for the effective
action we use an expansion in superspace (see Appendix \ref{sec:Superspace} for
 our conventions)\footnote{For the Fourier transformation we use the convention
$\partial_j\to ip_j$ with the notations $\vect p=(p_1, p_2)^T$ and
$p=\abs{\vect p}$ where there is no risk of misunderstanding.}
\begin{align}
\Gamma_k=&
-2\int d^2x\int dy\, d\bar y\;Z_k^2(\partial\parbar)\bar\Phi\Phi\\
\nonumber
&-2\int d^2x\int dy\;W_k(\Phi)-2\int d^2x\int d\bar y\;\bar W_k(\bar\Phi)
\\
=&
\int
\frac{d^2{p}}{4\pi^2}\left[Z_k^2(p^2)\left(2 p^2\phibar\phi+\psibar
i\slashed{\vect{p}}\psi-\frac12\Fbar F\right)\right.\label{eq:ansatz}\\
\nonumber
&\left.+\frac12FW'_k+\frac12\Fbar\,\Wbar'_k
+\psibar\left(W''_kP_++\Wbar''_kP_-\right)\psi\right].
\end{align}  
In contrast to the usual supercovariant derivative expansion we have included 
those combinations of the supercovariant derivatives that merely reduce to
spacetime derivatives. A momentum dependence in $W_k$ is irrelevant as found in section \ref{sec:NonRenTheorem}.
An arbitrary K{\"a}hler potential ($K(\bar\Phi,\Phi)$ integrated over the whole superspace) is not taken into account here, since we expect only a small influence for the renormalized mass.
Another contribution neglected in this truncation comes from the terms of higher than quadratic order in the  auxiliary field and the corresponding supersymmetric partner terms, denoted as auxiliary field potential. 
In the following
we will only work with real and imaginary part $\phi_1,\phi_2,F_1,F_2$.

For this scale dependent effective action the auxiliary fields obey the
equations of motion $F={\Wbar'_k(\phi)}/{Z^2_k}$
{and} $\Fbar={W'_k(\phi)}/{Z^2_k}$.
This leads to the on-shell action
\begin{multline}\label{eq:onshell}
\Gamma_k^{\rm on}=\int\frac{d^2 p}{4\pi^2}\left[
	\frac{1}2Z_k^2(p^2) p^2
	\phi\phibar+\frac12\frac{\abs{W'_k(\phi)}^2}{Z_k^2(p^2)}\right.\\
	\left.+iZ_k^2(p^2)\psibar
\slashed{\vect{p}}\psi+\psibar(W''_kP_++\Wbar''_kP_-)\psi	\right].	
\end{multline}

\subsection{Supersymmetric regulator}
Supersymmetry is preserved if we shift the mass by a momentum-dependent
infrared regulator\footnote{The regulator function is multiplied with the wave
function renormalization to ensure reparametrization invariance of the flow
equation.}, $m\to m+{Z_k^2\cdot r_1(k,p^2)}$ or multiply
the wave function renormalization by a momentum-dependent regulator function,
$Z_k^2\to {Z_k^2\cdot r_2(k,p^2)}$.
Such regulators are the same as  the ones used in the previous models
\cite{Synatschke:2010ub,Synatschke:2008pv,Synatschke:2009nm,Gies:2009az}.
To get a regularized path integral $R_k$ is included in terms of the cutoff
action $\Delta S_k$. It reads in a matrix notation 
 \begin{align}
	\Delta S_k=\frac12\int
	\frac{d^2{p}}{4\pi^2}\;\bar\Psi\,
		Z_k^2 R_k^T\,\Psi^T
\label{eq:CutoffAction}
\end{align}
with $\Psi=(\phi_1\;\phi_2\;F_1\;F_2\;\psi(-\vect p)^T\;\psibar(\vect p))$ and  
\begin{equation}
  \begin{split}
&R_k=\begin{pmatrix}
    R_k^{B}&0\\0&R_k^F
    \end{pmatrix}
\;\text{with}\; R_k^B=
\begin{pmatrix}
	p^2r_2\cdot\mathds{1}&r_1\cdot\sigma_3\\
	r_1\cdot\sigma_3&-r_2\cdot\mathds{1}
\end{pmatrix}\\
&\text{and}\; R_k^F=
\begin{pmatrix}
0&i\slashed{\vect p}\cdot r_2-r_1\cdot\mathds{1}\\
i\slashed{\vect p}\cdot r_2+r_1\cdot\mathds{1}&0
\end{pmatrix}.
\end{split}
\end{equation}
With these regulators at hand we can proceed to calculate the flow equation. Inserting ansatz
\eqref{eq:ansatz} in the flow equation \eqref{eq:lpa1}, the propagator can be
calculated along the lines described in \cite{Gies:2001nw}: The fluctuation matrix $\Gamma_k^{(2)}+R_k$ is
decomposed into the propagator $\Gamma_0^{(2)}+R_k$ including the regulator
functions and a part $\Delta \Gamma_k$ containing all field dependencies.  The
flow equation \eqref{eq:lpa1} is expanded in the number of fields, see
Appendix~\ref{sec:DerivationFlowEquation} for the expansion and the explicit
matrices.

\subsection{Flow equation for the superpotential -- The nonrenormalization
theorem}
\label{sec:NonRenTheorem}

The quantity at leading order in the supercovariant derivative expansion is the
scale dependent superpotential. We obtain the flow equation  by projecting onto
the terms linear in the auxiliary fields. We can choose either the real or imaginary part
of the auxiliary field as they are bound to give the same
results due to supersymmetry. The superpotential
$W(\phi)=u(\phi_1,\phi_2)+iv(\phi_1,\phi_2)$ is a holomorphic function of
$\phi_1$ and $\phi_2$, and therefore its real and imaginary part obey the Cauchy-Riemann differential equations \begin{equation}\frac{\partial u}{\partial\phi_1}=\frac{\partial v}{\partial\phi_2},\quad \frac{\partial u}{\partial\phi_2}=-\frac{\partial
v}{\partial\phi_1}.
\end{equation} 
Using these equations we find for the flow
equations of the superpotential
  \begin{equation}
  \partial_k u_k=0,\;\partial_k v_k=0\;
\Rightarrow\;\partial_kW_k=\partial_k\Wbar_k=0.
\end{equation}
This means that the superpotential remains unchanged during the RG flow.
The K{\"a}hler potential does therefore not influence the flow of the superpotential, as found in \cite{MirrorSymmetry}. 
Even the nontrivial momentum dependence considered here does not change this result.
The nonrenormalization theorem is verified by the flow equations in the present truncation.
This result is similar to the proofs in four dimensions discussed in
\cite{Sonoda:2009df} and \cite{Rosten:2008ih}.

As the flow vanishes at leading order, the first quantity with a nonvanishing
flow is the wave function renormalization which is a term at next-to-leading order in our truncation. It
will turn out later that the momentum dependence is important for the renormalized mass 
(cf.~Sec.~\ref{sec:results}) therefore we already include it in ansatz
\eqref{eq:ansatz}. 

\subsection{Momentum-dependent flow equation for the wave function
renormalization}

The flow equation for the wave function renormalization can be obtained from a
projection onto the terms quadratic in the auxiliary fields. 
It is derived in
Appendix~\ref{sec:DerivationFlowEquation} and  reads
\begin{multline}
\partial_kZ_k^2(p)=-8 g^2\int\frac{d^2 q}{4\pi^2}\frac{ h(\vect p
-\vect q)  h(\vect q)}
  {v(\vect q)^2 v(\vect p-\vect q)^2 }\times\\ 
 \left[\partial_kR_1(\vect q-\vect
  p) M(\vect p-\vect
   q) v(\vect q)+\partial_kR_1(\vect q) M(\vect q)
  v(\vect p-\vect q)\right]\\
   +4 g^2 \int\frac{d^2 q}{4\pi^2}\frac{h(\vect p -\vect q)
   \partial_kR_2(\vect q) u(\vect q)
  v(\vect p -\vect q)}
  {v(\vect q)^2 v(\vect p-\vect q)^2   }  \\
  +4 g^2 \int\frac{d^2 q}{4\pi^2}\frac{h(\vect q) \partial_kR_2(\vect q-\vect p)
  v(\vect q)
  u(\vect p-\vect q)}
  {v(\vect q)^2 v(\vect p-\vect q)^2   }  
\label{eq:FlowWFR}
\end{multline}
with the abbreviations (recall that $\abs{\vect q}=q$)
\begin{align}
h(\vect q)=&\left(r_2\left( q\right)+1\right) Z^2_k\left(
q\right),\\ \nonumber
M(\vect q)=&\mbar+r_1( q)Z^2_k\left( q\right),\;
R_i(\vect q)=r_i\left( q\right)Z_k^2\left( q\right),\\ \nonumber
u(\vect q)=&M(\vect  q)^2- q^2h^2(\vect q),\;
v(\vect q)=M(\vect q)^2+q^2h^2(\vect q).
\end{align} 
Here we are dealing with a UV-finite theory and  therefore it is sufficient
to use the simple, masslike infrared regulator 
\begin{align}
	r_1(k,p^2)=k\quad\text{and}\quad
	r_2(k,p^2)=0.
\end{align}
After a shift in the integration variables in the second part of the integral \eqref{eq:FlowWFR}
the flow equation simplifies to
\begin{multline}
  \partial_kZ_k^2(p)=-16 g^2\int\frac{d^2 q}{4\pi^2}\frac{k
Z^2_k\left( q\right)+\mbar }
{N(\vect q)^2N(\vect p-\vect q)}\times\\
Z^2_k\left(q\right) Z^2_k\left(\abs{\vect p -\vect q}\right)\partial_k
\left(kZ_k^2\left( q\right)\right)   \label{eq:FlowEq},
\end{multline}
where we have introduced the abbreviation
\begin{equation}
N(\vect q)=\left(q^2 Z^4_k\left( q\right)+(kZ^2_k\left(
q\right)+\mbar)^2\right).
\end{equation} 

 In order to deal with the
partial differential equation we have developed a
(parallelizable)  numerical toolbox called FlowPy.   
In the next section we  present the main ideas behind our numerical setup
to solve the momentum-dependent flow equation. A detailed presentation of FlowPy
is deferred to a future paper \cite{tbp}.

\section{Numerical Setup}

\label{sec:NumRes}

Structurally, the flow equation to be solved numerically is
of the form
\begin{multline}
\partial_k z(k;p) = \int d^n q \,
I\left[k;p;q;z(k;f_1(p,q));\right.\\\left.z(k;f_2(p,q));\ldots;z(k;f_n(p,q))\right]\, ,
\end{multline}
where the external momentum $p$ is treated with a discretized grid. 
While the integrand~$I$ may actually also be a function of~$\partial_k
z(k;p)$, and hence the integral flow equation be given in implicit
form only, numerical results suggest that, at least in the
model studied here, $\partial_k z(k;p)$~should be sufficiently
small to allow an iterative approach, where the integrand is evaluated
first under the assumption~$\partial_k z(k;p)=0$, and the result is
then used to re-evaluate the integrand with a better approximation to
$\partial_k z(k;p)$ until convergence is reached. Apart from this conceptual
issue, the technology to deal with an evolution equation of this kind is readily available
in an accessible form via the SciPy ``Scientific Python''
extension~\cite{SciPy} to the Python~\cite{Python} programming
language. The details of the numerical strategy are described in Appendix~\ref{sec:CompStrat}.

\begin{figure} 
\includegraphics[width=.48\textwidth]{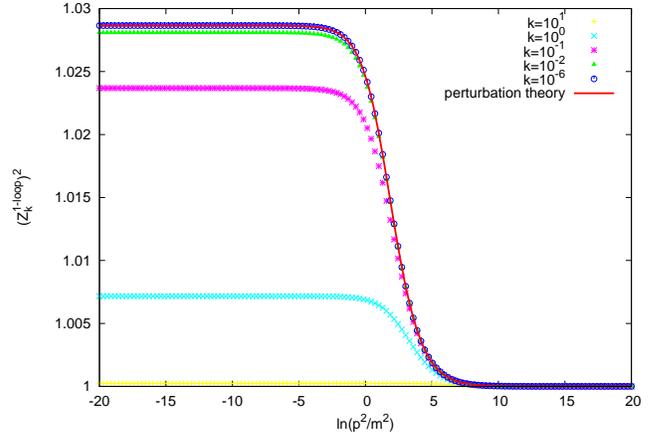}	
\caption{Perturbative flow for the parameters
$\lambda={\gbar}/{\mbar}=0.3$ and $\mbar=1$ . The solid line
is the plot of Eq.~\eqref{eq:OneLoop}.  \label{fig:CompPertTheory}}
\end{figure}

As a test for the numerical approximation and the abilities 
of FlowPy, we solve the flow of the perturbative wave function
renormalization. It is inferred  by setting $Z_k(q)$ to its classical value
$Z_k(q)\equiv1$ on the right hand side of Eq.~\eqref{eq:FlowEq}. The
perturbative flow with $60$ discretization points is shown in Fig.~\ref{fig:CompPertTheory}
for different values of $k$.

It is possible to calculate the perturbative expression for $Z^2_{\rm
1-loop}(p)$ analytically from the perturbative flow equation by
performing the $k$-integral using $\lim_{k\to\infty}r_1,r_2\to\infty$ and $\lim_{k\to0}r_1,r_2\to0$. This yields
\begin{align}
\nonumber
  Z^2_{\rm 1-loop}
=&1+\frac{\gbar^2}{\pi^2}\int \;
\frac{d^2{q}}{(\mbar^2+ q^2)(\mbar^2+\abs{\vect q-\vect p}^2)}\displaybreak[0]
\nonumber\\
=&1+4\gbar^2\frac{\mathrm{artanh}\left({p({4\mbar^2+p^2})^{-1/2}}\right)}{\pi
p\sqrt{4\mbar^2+p^2}}\,,
\label{eq:OneLoop}
\end{align}
which is shown as a solid line in Fig.~\ref{fig:CompPertTheory}. This shows
that possible errors in the numerical calculation of the wave function renormalization with FlowPy due to discretization, interpolation 
and the boundary condition $Z_k(q\to\infty)=1$ are under
control. We will consider $Z_k$ and the renormalized masses obtained from it as exact in the employed truncation.

In the next section we will determine renormalized masses from the
nonperturbative wave function renormalization with full momentum dependence
calculated with FlowPy. 

\section{The renormalized mass}
\label{sec:results}
The analytic continuation of the bosonic propagator
\begin{equation}
G_{\rm bos}(p)=\frac{1}{p^2+\mbar^2+\Sigma(p,\mbar,g)}
\label{eq:polemass}
\end{equation} 
has a pole which defines the renormalized mass. 
Since the bare mass
$\mbar$  is a parameter of the superpotential \eqref{eq:spot} it is not changed
during the flow. $\Sigma$ is the self-energy. As expected from a
supersymmetric theory, the pole of the fermionic propagator leads to the same
renormalized mass as the bosonic propagator.

 The Fourier transformation of $G_{\rm bos}(p)$ yields the correlator 
 \begin{equation}
    C_{\rm bos}(x_1)=\int\frac{dp}{2\pi}G(p_1,0)e^{ip_1x_1}. 
 \end{equation}
The renormalized mass can be obtained from the long range exponential decay of this quantity
 and is in the following denoted
 as \emph{correlator} mass  $\mcor$. One can also define  a renormalized mass,
 which we denote as \emph{propagator} mass, 
through $m_{\rm prop}^2=(G_{\rm bos}(p))^{-1}|_{p=0}$ .

To compare the renormalized masses from the FRG with the results of the lattice
simulation \cite{Kastner:2008zc} we consider the masses of the particles in the on-shell
theory. In our truncation the bosonic propagator from the on-shell
action \eqref{eq:onshell} reads in the infrared limit
\begin{equation}
	G_{\rm bos}^{\rm NLO}(p)=\frac{1}{p^2Z_{k\to0}^2(p^2)+\mbar^2/Z_{k\to0}^2(p^2)}
\label{eq:PropagatorBoson},
\end{equation}
and the fermionic propagator reads
\begin{equation}
	G_{\rm ferm}^{\rm NLO}(p)=\frac{\slashed{p}}{p^2Z^4_{k\to0}(p^2)+\mbar^2}
	\label{eq:PropagatorFermion}.
\end{equation}
Both propagators have the same poles and therefore lead to the same renormalized masses for bosons and fermions.

For the propagator mass the fields in the on-shell action have to be
rescaled with the wave function renormalization such that the kinetic term
is of the canonical form. Neglecting the momentum dependence in the wave
function renormalization we obtain
\begin{equation}
  m_{\rm prop}=\frac{m}{Z_{k\to0}^2(p=0)}.\label{eq:Propagatormasse}
\end{equation}
For a 
small self-energy $\Sigma$ a  comparison between Eq. \eqref{eq:polemass} and
\eqref{eq:PropagatorBoson} leads  to the approximate relation
\begin{equation} 
Z^2_{k\to0}(p)=1+\frac{\Sigma(p,\mbar,g)}{p^2-\mbar^2}.
\label{eq:Polarisation}
\end{equation}  
A numerical calculation can provide $Z_k^2(p)$ only for real $p$ and its
analytic continuation cannot be determined straightforwardly.
Instead we consider the discrete Fourier transformation of $G_{\rm bos}^{\rm
NLO}(p)$ with  momenta $p=\{0,{2\pi}/{aN},\ldots,{2\pi (N-1)}/{a N} \}$ on
the interval $x\in[0,aN=L]$. For  distances much smaller than $L$ this should
approximate $C_{\rm bos}^{\rm NLO}(x)$ in a well-defined way.  More precisely,
instead of the exponential decay  one gets  the long distance behavior
\begin{equation}
C_{a,m_{\rm cor}}(x_1)\propto\cosh(\mcor(x_1-L/2))\, 
\end{equation}
after the integration over the spatial direction.
The mass can be determined from a fit to this function, as it is done in lattice
simulations. The details of this procedure can be found in
Appendix~\ref{sec:DetRenMass}.

With the analytic result \eqref{eq:OneLoop} for $Z^2_{\rm 1-loop}$ at hand we
can calculate the poles of $G_{\rm bos}^{\rm NLO}(p)$ and obtain a perturbative
approximation of $\mcor$.Note that this analytic solution of the perturbative flow together with eq.
\eqref{eq:Polarisation} leads to the same result as a one-loop \emph{on-shell} 
calculation of the polarization $\Sigma$ (cf.~Appendix~\ref{sec:PertTheory}). 
Expanding the pole of the propagator \eqref{eq:polemass} 
to first order in the dimensionless parameter
$\lambda^2=g^2/m^2$ leads to the one-loop approximation of the renormalized mass
\begin{equation}
(\mcor^{\rm 1-loop})^2=\mbar^2\left(1-\frac{4}{\sqrt{27}}\lambda^2
	+\mathcal O\left(\lambda^4\right)\right)\, .
\label{eq:OneLoopCorrMass} 
\end{equation}

\subsection{Weak couplings}
\label{sec:weak}

Let us start with an investigation of the weak coupling sector which is defined
as $\lambda<0.3$, where perturbation theory provides an excellent cross-check to
establish the correctness of  our ansatz and the errors in the numerical approximation.
 
The
bare mass in the lattice simulations \cite{Kastner:2008zc} is taken to be $m=15$. Concerning
the units of the mass note the following: In the lattice calculation, the mass
is measured in units of the box size, i.\,e. the physical volume of the lattice
simulation. Similarly, everything can be reformulated in terms of the dimensionless 
ratio of bare and renormalized mass.

 For the numerical treatment of
Eq.~\eqref{eq:FlowEq} we have to use dimensionless quantities. Because of the
nonrenormalization theorem the bare mass quantities in the
superpotential enter in the flow equation only as parameters. Rescaling
the dimensionful quantities  with the bare mass sets the scale in this model. 
We have set this scale to $m=1$. To get the same units as in the lattice
simulations the resulting renormalized mass is multiplied with 15.

 \begin{figure}
 	\includegraphics[width=.45\textwidth]{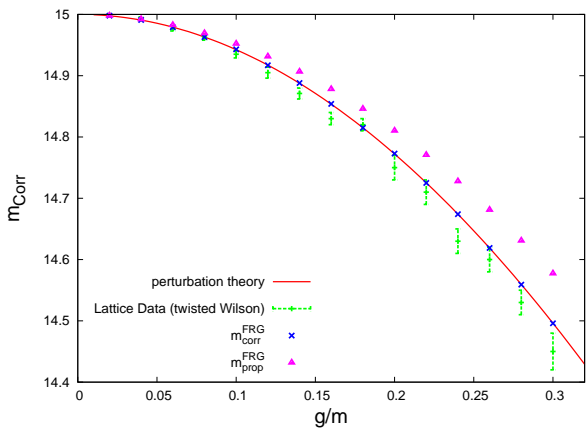} 
 	\caption{Comparison between lattice data taken from \cite{Kastner:2008zc} and
 	our results for the correlator mass $m_{\rm corr}^{\rm FRG}$ \emph{with}
 	momentum dependence and $m_{\rm prop}^{\rm FRG}$ without momentum dependence
 	in the weak coupling regime\label{fig:CompLatWeak}}
 \end{figure}

\begin{table}
\begin{small} 
\caption{Renormalized masses obtained with the flow equation
with and without momentum dependence ($\mcor^{\rm FRG}$ and $m_{\rm prop}^{\rm
FRG}$) as well as  lattice data $\mcor^{\rm lattice}$ from a continuum
extrapolation
\cite{Kastner:2008zc} in the weak coupling regime.\label{tab:ComparisionLatticeWeakCoupling} }
\begin{ruledtabular}
\begin{tabular}{ccccc}
$\lambda$ &$\mcor^{\rm FRG}$ & $m_{\rm prop}^{\rm FRG}$& $\mcor^{\rm
lattice}$\\\hline 0.02&14.998&14.998&  14.999(1)\\
0.04&14.991&14.992&  14.993(3)\\
0.06&14.979&14.983&  14.977(4)	\\
0.08&14.963&14.970&  14.963(5)	\\
0.10&14.943&14.952&  14.935(6)  \\	
0.12&14.917&14.931&  14.905(9)\\
0.14&14.888&14.907&  14.871(9)	\\
0.16&14.854&14.878&  14.83(1)\\
0.18&14.815&14.846&  14.82(1) \\
0.20&14.773&14.810&  14.75(2) 	\\
0.22&14.674&14.771&  14.71(2)	\\ 
0.24&14.674&14.728&  14.63(2)	\\
0.26&14.619&14.681&  14.60(2)   \\
0.28&14.559&14.631&  14.53(2)\\	
0.30&14.496&14.578&  14.45(3)    \\    
\end{tabular}
\end{ruledtabular}
\end{small}

\end{table}

The correlator masses  in the weak coupling regime are calculated with the
momentum-dependent wave function renormalization from the flow equation
\eqref{eq:FlowEq} solved with FlowPy. The technical details of the determination of the
correlator masses are described in Appendix~\ref{sec:DetRenMass}. The results are shown in
the second column of Table \ref{tab:ComparisionLatticeWeakCoupling}.  The values in the
fourth column are taken from a Monte-Carlo  simulation on  the lattice
\cite{Kastner:2008zc}.   We discuss the lattice results  further in
Sec.~\ref{sec:intermediate}. For the time being it suffices to note the
agreement of lattice and perturbative results within  the statistical errors. Hence perturbation theory already provides a 
good check for our results.

In Fig.~\ref{fig:CompLatWeak} we show the correlator masses from the flow
equation, the lattice simulation and the one-loop result
\eqref{eq:OneLoopCorrMass} for $\mcor$. The masses calculated from the flow
equation agree very well with perturbation theory and with the results from
lattice simulations. This can be quantified by comparing the  correction to
the bare mass $\Delta m_{\rm corr}=m-\mcor$.  We
find $\Delta \mcor^{\rm FRG}/\Delta \mcor^{\rm lattice}\simeq0.95$. 
Taking into account the statistical error of the lattice data no significant 
difference to the FRG results can be found. 

We conclude that in the weak coupling regime the truncation of the
flow equation  with full momentum dependence
suffices  to capture the main aspects of the model.
Higher-order-operators, which yield an auxiliary field effective potential, 
have, as expected, little influence.  

To investigate the influence of the momentum dependence in the wave function
renormalization, we calculate the propagator mass \eqref{eq:Propagatormasse}.
The results are shown in the third column of
Tab.~\ref{tab:ComparisionLatticeWeakCoupling} and in
Fig.~\ref{fig:CompLatWeak}. A comparison between the propagator mass and the
correlator mass from the lattice calculation yields $\Delta
m_{\rm prop}^{\rm FRG}/\Delta \mcor^{\rm lattice}\simeq0.75$. Already in the
weak coupling regime it is necessary to include the momentum dependence in
order to determine the corrections to the renormalized mass with satisfying
accuracy.

\subsection{Intermediate couplings}
\label{sec:intermediate}
At intermediate couplings $0.3\leq\lambda\leq1$ a significant deviation of
our numerical results from perturbation theory can be observed.  In this regime the
perturbative calculations can no longer provide a reliable  test for the
numerical results and we have to rely on lattice calculations. 
 In a supersymmetric theory their  result must,
however, be considered with care as a lattice formulation of supersymmetry still
poses difficulties 
\cite{Bergner:2009vg}. A
common approach is  to implement only a  part of the supersymmetry
which allows  to recover the complete symmetry in the continuum limit in many
cases.

\begin{figure}
	\includegraphics[width=.45\textwidth]{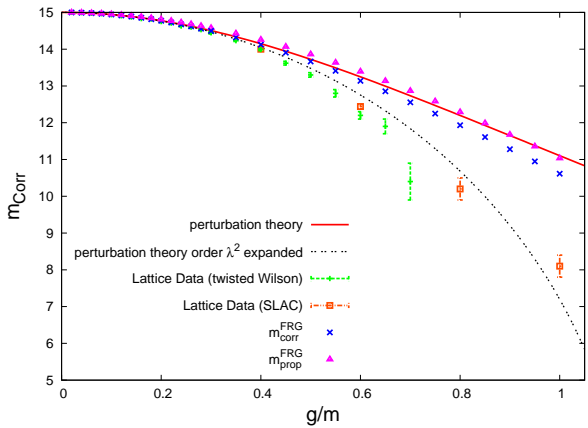} 
	\caption{Comparison between lattice data taken from \cite{Kastner:2008zc} and
 	our results for the correlator mass $m_{\rm corr}^{\rm FRG}$ \emph{with}
 	momentum dependence and $m_{\rm prop}^{\rm FRG}$ without momentum dependence
 	in the intermediate coupling regime \label{fig:CompLat}}
\end{figure}

In the present model there are further complications for a lattice formulation,
especially in the intermediate coupling sector \cite{Kastner:2008zc}.  The
considered discretizations are invariant under half of the supersymmetry. 
They suffer, however, from the dominance of a contribution to the
action   that is a mere discretization of a surface term at larger couplings.
In general the correct continuum limit can only be obtained with unrealistically
high numerical effort. The relevance of  this effect depends on the coupling
strength and on the specific discretization.  For intermediate couplings the nonlocal \emph{SLAC} 
discretization and the \emph{Twisted Wilson} discretization
provides the most reliable results
(cf.~\cite{Kastner:2008zc} for details). The renormalized masses of these
discretizations are used for a comparison with our results. They are shown in the third and fourth column of Table 
\ref{tab:ComparisionLattice} \footnote{All lattice results are extrapolated to
the continuum.} and displayed in Fig.~\ref{fig:CompLat} (boxes with error bars) together with
the order $\lambda^2$ expanded result \eqref{eq:OneLoopCorrMass} for $\mcor$
(dashed line).

Note that the spontaneous breaking of the $\mathbb{Z}_2$ symmetry introduces also finite volume effects in the lattice simulations.
Although there are well known prescriptions to implement these properties of
the theory in lattice simulations, they still lead to additional complications
\cite{Wozar}.

\begin{table}
\caption{Masses obtained with the flow equation
with and without momentum dependence $\mcor^{\rm FRG}$ and $m_{\rm prop}^{\rm
FRG}$) as well as  lattice data 
\cite{Kastner:2008zc} in the regime with intermediate  couplings.\label{tab:ComparisionLattice}}
\begin{small}
\begin{ruledtabular}
\begin{tabular}{ccccc}
$\lambda$&$\mcor^{\rm FRG}$ & $m_{\rm prop}^{\rm FRG}$&  
tw. Wilson& SLAC imp.\\\hline 
0.35&14.321&14.428&14.23(2)&		 \\ 
0.40&14.123&14.259& 13.99(3)&14.00(1)\\
0.45&13.905&14.069& 13.62(5)&		 \\  
0.50&13.666&13.861& 13.30(6)&		 \\ 
0.55&13.411&13.636& 12.8(1) &		 \\
0.60&13.138&13.394& 12.2(1) &12.44(6)\\
0.65&12.854&13.137& 11.9(2) & 		 \\
0.70&12.556&12.866&	10.4(5) &		 \\
0.75&12.248&12.583&			&		 \\
0.80&11.932&12.290& 		&10.2(3) \\
0.85&11.609&11.987&			&		 \\
0.90&11.280&11.676&			&		 \\
0.95&10.948&11.358&			&	  	 \\
1.00&10.613&11.036&  		&8.1(3)\footnote{C. Wozar, private communication}   
\end{tabular}
\end{ruledtabular}
\end{small}

\end{table}

The correlator masses determined from the FRG  are shown in the second
column of Tab.~\ref{tab:ComparisionLattice} and displayed in Fig.~\ref{fig:CompLat} (crosses). Additionally the perturbative result
for the renormalized mass is shown (solid line), which is determined from the
pole of the propagator \eqref{eq:polemass} with the polarization calculated in Appendix
\ref{sec:PertTheory}\footnote{The perturbative results in this regime have to be interpreted with care. The dashed line is an expansion of the on-shell one-loop pole mass to order $\mathcal O(\lambda^2)$ whereas the solid line is the result of an off-shell calculation.  Both results agree up to order $\lambda^2$ but perturbation theory can no longer be trusted in the regime with intermediate coupling strength. The on-shell calculation has to fail at large values of $\lambda^2$ because otherwise the renormalized masses will become negative. To preserve supersymmetry in the RG flow the FRG uses an off-shell formulation. Therefore 
it is not unexpected that it is close to the off-shell perturbation theory.}. 

Although the corrections  from the wave function renormalization
with full momentum dependence to the bare mass capture some of the
quantum effects, they do not account for \emph{all} the nonperturbative effects
present in this model. To quantify this, we 
compare these corrections
to the
corrections found in lattice calculations. This yields results between $\Delta
\mcor^{\rm FRG}/\Delta \mcor^{\rm lattice}\simeq0.9$  for $\lambda=0.35$ and
$\Delta \mcor^{\rm FRG}/\Delta \mcor^{\rm lattice}\simeq0.65$ for
$\lambda=1.0$. The fact that the wave function renormalization accounts for less
of the quantum corrections as the coupling grows is  due to the growing
influence of higher-order operators especially the auxiliary field
potential.  In the present
truncation we have only considered terms that are at most   quadratic in the
auxiliary field and have neglected back reactions from  a potential for the auxiliary field. As can be seen from
a diagrammatic expansion of the flow equation, terms up to order $F_i^4$ 
directly modify the flow equation  for the wave function renormalization, 
which is proportional to $F_i^2$.   It is known from our previous
investigations of scalar supersymmetric  models \cite{Synatschke:2008pv} that
the influence of higher order operators grows with the strength of the
couplings.  A truncation that goes beyond the momentum-dependent wave
function renormalization has to be considered to improve the results in the
regime with intermediate couplings.

The results for the propagator mass are shown in the third column  of
Tab.~\ref{tab:ComparisionLattice} and in Fig.~\ref{fig:CompLat} (triangles).
Compared to the lattice results we find  $\Delta m_{\rm prop}^{\rm FRG}/\Delta
\mcor^{\rm lattice}\simeq0.75$ for $\lambda=0.35$ and $\Delta m_{\rm prop}^{\rm
FRG}/\Delta \mcor^{\rm lattice}\simeq0.6$ for $\lambda=1.0$. The improvement
due to the momentum dependence in $Z_k^2$ is not as pronounced as it is in the
weak coupling regime. 

\section{Conclusions}

In this paper we have applied the functional
renormalization group to the $\N=(2,2)$ Wess-Zumino model in two dimensions.
The model is UV-finite which allows a direct comparison to results from lattice
simulation. 

The first quantity to be calculated in a supercovariant derivative expansion is
the superpotential. It is well known from the nonrenormalization theorem that
it does not receive quantum corrections. In the language of the FRG the
nonrenormalization theorem is recovered in a very simple form, namely that
the superpotential has a vanishing flow equation. The proof only
uses the fact that the superpotential is a holomorphic function and therefore
the Cauchy-Riemann differential equations for its real and imaginary parts hold.

Hence the first term in the expansion that receives a correction from renormalization 
is the wave function renormalization.
It leads to the renormalization of the bare mass in the on-shell theory,
with the renormalized mass defined as the pole of the propagator in the complex
plane. We have calculated the renormalized mass with and without momentum
dependence in the wave function renormalization.

In order to benchmark our results we use lattice calculations. In the weak coupling regime the results for the
renormalized mass calculated without the full momentum dependence capture only 75\% of the quantum
corrections to the bare mass whereas 95\% of the corrections are captured if
the full momentum dependence in the wave function renormalization is taken into
account. 
 This leads to the conclusion that the momentum dependence of  wave function
 renormalization dominates in this regime. Higher-order-operators  only have a small influence.

For intermediate couplings the picture changes. We have investigated the
complete contribution to the flow from the momentum-dependent wave
function renormalization. Our findings are that in this truncation only
65\% of the quantum corrections to the bare mass determined in the lattice simulations are captured for the largest coupling
considered in this paper. Without momentum dependence 60\% of the corrections
are generated. This leads to the conclusion  that  in the  regime with
intermediate couplings the  momentum dependence  in the wave function
renormalization does not include all important contributions to the renormalized
mass. Instead, the quantum corrections generated by 
higher-order-operators which lead to an auxiliary field potential are expected to be relevant for the renormalized mass. 
They have to be included in order to reduce the deviations between the results from lattice calculations and the FRG.
The calculation of these contributions as well as contributions from the
K{\"a}hler potential remains an interesting challenge for future work.

Although there have been great improvements  in the simulations of the model on
the lattice they still suffer from finite size effects and the finite lattice
spacing, which leads to a breaking of supersymmetry\footnote{ In contrast to common lattice formulations the basic requirements of locality and reflection positivity are broken in the current simulations of this theory. This was done to reduce the unavoidable breaking of supersymmetry on the lattice.}. 
An interesting application of the  FRG is an  analysis that
includes finite volume effects which could help to estimate the influence of the finite size. This can allow  one to separate it from the discretization errors. 

The  analysis presented in this paper can easily be applied to the $\N=1$
Wess-Zumino model in four dimensions from which the two-dimensional $\N=(2,2)$ 
model  is derived.  Especially the nonrenormalization theorem for the
superpotential emerges in the same way. In both models the momentum-dependent
wave  function renormalization is the first relevant contribution  in the
covariant derivative expansion and the flow equations differ only in the
measure of integration. As we have found in the present model the effective
potential for the auxiliary field  is expected to dominate  the quantum effects as the strength of the  coupling
constant grows.

For the treatment of the partial differential equation we have developed 
FlowPy, a numerical toolbox. It can be applied in quite generic situations to solve the FRG equations 
and to calculate contributions such as the full momentum dependence of vertices.
We hope that numerical software like FlowPy will  
help to obtain better predictions from  FRG calculations. We would like to
add that it can also be applied for the calculation of an 
arbitrary potential $V(\phi)$ instead of the full momentum dependence $Z(p)$. 
We plan to make FlowPy available soon~\cite{tbp}.   

\acknowledgments{We thank J.~Braun, H.~Gies, A.~Wipf
and C.~Wozar for helpful discussions and valuable comments on the
manuscript as well as S.~Diehl and U.~Theis for helpful discussions.  FS
acknowledges support by the Studienstiftung des deutschen Volkes. This work has been supported by the DFG-Research Training Group 
''Quantum-and Gravitational Fields'' GRK 1523/1.}
 
\appendix

\section{Superspace formulation}
\label{sec:Superspace}

The superspace formulation  is constructed from the supersymmetry
transformations. The chiral and antichiral superfields can be obtained from the
lowest component by acting on it with the exponentiated supersymmetry transformations
\begin{align} \nonumber
\Phi&(z,\zbar,\alpha,\bar\alpha)=\exp(-\delta_\alpha)\phi(z,\zbar)
	=\sum_{n=0}^4\frac1{n!}(-\delta_\alpha)^n\phi(z,\zbar)	
\\ \nonumber
=&\phi(u,\ubar)
	-\psibar_1(u,\ubar)\alpha_1
	-\bar\alpha_1\psi_1(u,\ubar)
	-\frac{F(u,\ubar)}2 \bar\alpha_1\alpha_1
\displaybreak[0]
	\\ 
	\nonumber
\bar\Phi&(z,\zbar,\alpha,\bar\alpha)=\exp(-\delta_\alpha)\phibar(z,\zbar) 
	=\sum_{n=0}^4\frac1{n!}(-\delta_\alpha)^n\phibar(z,\zbar)\\ 
	=&\phibar(u,\ubar)
-\psibar_2(u,\ubar)\alpha_2-\bar\alpha_2\psi_2(u,\ubar)
-\frac{\bar F(u,\ubar)}2\bar\alpha_2\alpha_2
\end{align}
 with the chiral variables $u=z-\frac12\bar\alpha_2\alpha_1$ and
 $\ubar=\zbar+\frac12\bar\alpha_1\alpha_2$.
The supercharges are 
 \begin{equation}
    \begin{split}
Q_1=-\frac{\partial}{\partial\bar\alpha_1}+\frac12\alpha_2\parbar,\quad
\bar Q_1=\frac{\partial}{\partial\alpha_1}-\frac12\bar\alpha_2\partial
\\
Q_2=-\frac{\partial}{\partial\bar\alpha_2}+\frac12\alpha_1\partial,\quad
\bar Q_2=\frac{\partial}{\partial\alpha_2}-\frac12\bar\alpha_1\parbar,
    \end{split}
 \end{equation}
 and the supercovariant derivatives read
 \begin{equation}
    \begin{split}
D_1=-\frac{\partial}{\partial\bar\alpha_1}-\frac12\alpha_2\parbar,\quad
\bar D_1=\frac{\partial}{\partial\alpha_1}+\frac12\bar\alpha_2\partial
\\
D_2=-\frac{\partial}{\partial\bar\alpha_2}-\frac12\alpha_1\partial,\quad
\bar D_2=\frac{\partial}{\partial\alpha_2}+\frac12\bar\alpha_1\parbar.    
    \end{split}
 \end{equation}
 The superfield obeys the (anti)chiral constraint
 \begin{equation}
    D_2\Phi=\bar D_2\Phi=0,\quad
    D_1\bar\Phi=\bar D_1\bar\Phi=0.
 \end{equation}
The supersymmetry transformations are generated by
\begin{equation}
\delta\Phi=(\bar\varepsilon Q+\bar Q\varepsilon)\Phi,\quad
\delta\bar\Phi=(\bar\varepsilon Q+\bar Q\varepsilon)\bar\Phi.
\end{equation}
The Lagrange density is given by 
 \begin{align}
\mathcal L=&\mathcal L_{\rm kin}+\mathcal L_{\rm pot}\\
=&-2\int dy\,d\bar y\; \bar\Phi\Phi
-2\int d y\; W(\Phi)
-2\int d\bar y\; \bar W(\bar\Phi)
\nonumber
 \end{align}
with
$dy\equiv d\bar\alpha_1 d\alpha_1$ {and} $d\bar y\equiv 
d\alpha_2d\bar\alpha_2$.
\section{Flow equation for the momentum-dependent wave
function renormalization}
\label{sec:DerivationFlowEquation}

To obtain the flow equations for the wave function renormalization we decompose the
second derivative of the effective action into a field independent part $\Gamma_0^{(2)}+R_k$  and  a field dependent part $\Delta\Gamma_k^{(2)}$ (in the following we drop the momentum dependence of the
regulators for simplicity of notation):
 \begin{multline}
(\Gamma_0^{(2)}+R_k)(\vect{q},\vect{q'})
	+\Delta\Gamma_k(\vect q,\vect q')
	\\
=
\begin{pmatrix}
A_0&0\\0&B_0
\end{pmatrix}\delta(\vect{q}-\vect{q'})
+
\begin{pmatrix}
\Delta A&\Delta C\\\Delta D&\Delta B
\end{pmatrix}
 \end{multline}
{with ($h=(1+r_2)Z_k^2(q)$, $M=(r_1Z_k^2(q)+\mbar )$)}
 \begin{equation}
    \begin{split}
A_0=
\begin{pmatrix}                   	
q^2h\cdot\mathds{1}
                   			&M\cdot\sigma_3 \\           
M\cdot\sigma_3&-h\cdot\mathds{1}
\end{pmatrix},
\;
 B_0=
i\slashed{\vect{q}}h+M\mathds{1}
    \end{split}
 \end{equation}
{and}
 \begin{equation}
    \begin{split}
&\Delta A=
2g\begin{pmatrix}     
   F_1 &-F_2 & \phi_1 &-\phi_2\\
 -F_2& -F_1 & -\phi_2 &-\phi_1 \\
 \phi_1& -\phi_2& 0 & 0  \\
  -\phi_2 & -\phi_1& 0 & 0  \\
\end{pmatrix}\left(\vect{q}+\vect{q'}\right)
,\\&
\Delta C=
2g\begin{pmatrix}
  \psibar_1&i\psibar_2\\
  \psibar_1&-i\psibar_2\\0&0\\0&0
  \end{pmatrix}\left(\vect{q}+\vect{q'}\right)\\
&
\Delta D=
2g\begin{pmatrix}
\psi_1&i\psi_1&  0&0\\
\psi_2&-i\psi_2  &0&0
  \end{pmatrix}\left(\vect{q}+\vect{q'}\right)
,\\& 
\Delta B=
2g\begin{pmatrix} 
\phi_1+i\phi_2 &
 0\\
 0 &\phi_1-i\phi_2
 \end{pmatrix}\left(\vect{q}+\vect{q'}\right).
    \end{split}
 \end{equation}
 The flow equation can then be expanded
\cite{Gies:2001nw} in 
 \begin{multline}
	\partial_t\Gamma_k=\frac12\tilde\partial_t\STr
	\left(({\Gamma_0^{(2)}+R_k})^{-1}\Delta\Gamma\right)\\-\frac14\tilde\partial_t\STr\left(({
\Gamma_0^{(2)}+R_k)^{-1}}\Delta\Gamma\right)^2+\ldots
 \end{multline}
with $\tilde\partial_t$ acting only on the regulator. $\STr$ denotes a trace in
field space as well as an integration in momentum space. 
The wave function renormalization is a term proportional to $F_i^2$ and can be
obtained from the second term in this expansion. 
To calculate this we
define
 \begin{align}
\nonumber
	M(\vect 
	q,\vect 
	q')\equiv&\int_{\vect 
	q''}({\Gamma_0^{(2)}+R_k})^{-1}(\vect q)\delta(\vect q+\vect 
	q'')\Delta\Gamma(\vect q'',\vect q')
		\\=&({\Gamma_0^{(2)}+R_k})^{-1}(\vect q)\Delta\Gamma(-\vect q,\vect 
		q')
 \end{align}
and the second term in the expansion reads 
\begin{align}
\nonumber
	&\tilde\partial_t\Str\int_{\vect q,\vect q'}M(\vect q,\vect q')M(\vect
	q',\vect q)\\
\nonumber	
	&=	\Str\int_{\vect	q,\vect q'}({\Gamma_0^{(2)}+R_k})^{-1}
		(\vect q)\partial_tR_k({\Gamma_0^{(2)}+R_k})^{-1}(\vect q) \\
\nonumber
&
	\phantom{\Str\int_{\vect	q,\vect q'}}	\times	
	\Delta\Gamma(-\vect	q,\vect q')
	({\Gamma_0^{(2)}+R_k})^{-1}(\vect q')
	\Delta\Gamma(-\vect q',\vect q)\\
\nonumber&
+\Str\int_{\vect	q,\vect q'}	
	({\Gamma_0^{(2)}+R_k})^{-1}(\vect q)
	\Delta\Gamma(-\vect	q,\vect	q')
	({\Gamma_0^{(2)}+R_k})^{-1}(\vect q')
	\\&\phantom{\Str\int_{\vect	q,\vect q'}}
	\times\partial_tR_k(\vect q')({\Gamma_0^{(2)}+R_k})^{-1}(\vect q')
			\Delta\Gamma(-\vect q',\vect q) 
 \end{align}
where $\Str$ denotes a trace in field space.
We take the functional
derivative with respect to $F_i(\vect p)$ and $F_i(-\vect p)$ and set all
fields to zero in order to project on the wave function renormalization
$Z_k(p^2)$.  This yields
\begin{align}
\nonumber
\partial_k&Z_k^2(p)=-8 \gbar^2\int\frac{d^2 q}{4\pi^2}\frac{ h(\vect p
-\vect q)  h(\vect q)}
  {v(\vect q)^2 v(\vect p-\vect q)^2 }\times\\ 
 \nonumber 
 &\left[\partial_kR_1(\vect q-\vect
  p) M(\vect p-\vect
   q) v(\vect q)+\partial_kR_1(\vect q) M(\vect q)
  v(\vect p-\vect q)\right]\\
\nonumber 
   &+4 g^2 \int\frac{d^2 q}{4\pi^2}\frac{h(\vect p -\vect q)
   \partial_kR_2(\vect q) u(\vect q)
  v(\vect p -\vect q)}
  {v(\vect q)^2 v(\vect p-\vect q)^2   }  \\
  &+4 g^2 \int\frac{d^2 q}{4\pi^2}\frac{h(\vect q) \partial_kR_2(\vect q-\vect p)
  v(\vect q)
  u(\vect p-\vect q)}
  {v(\vect q)^2 v(\vect p-\vect q)^2   }  
\end{align}
with the abbreviations 
\begin{align}
\nonumber
h(\vect q)=&\left(r_2\left(q\right)+1\right) Z^2_k\left(q\right),\\
M(\vect q)=&\mbar+r_1(q)Z^2_k\left(q\right),\;
R_i(\vect q)=r_i\left(q\right)Z_k^2\left(q\right),\\
\nonumber
u(\vect q)=&M(\vect q)^2-q^2h^2(\vect q),\;
v(\vect q)=M(\vect q)^2+q^2h^2(\vect q)
\end{align} 

\section{The computational strategy}
\label{sec:CompStrat}
To solve  flow equations depending on an external momentum the computational
strategy is as follows:
\begin{itemize}[label=$\bullet$, leftmargin=*]
\item Mapping the $k\to 0$ flow to a forward time evolution problem
  by introducing $\tilde k=-k$.

\item Discretization of the problem by approximating~$z(k;p)$ with an
 interpolation function that is determined by~$N$ support points.
 These are best chosen to be equidistant on a logarithmic scale,
 accounting for the expectation that a wide range of scales should
 contribute in a comparable way to the integral.

\item 
For now, we have been using interpolation in conjunction with the
two-dimensional integration function \texttt{scipy.integrate.dblquad()} from
Scientific Python. However, this ad-hoc approach should allow great
efficiency improvements by instead considering a tighter integration
of numerical quadrature with adaptive discretization. In particular,
it should then also be possible to also pass on information about the
discretized flow equation's Jacobian matrix to the numerical ODE
integrator. There hence is considerable potential for further
efficiency improvements of numerical RG flow code. 

\item Solving the resulting ordinary differential equation for the
 values of $z(k;p)$ at the support points with
 SciPy's~\texttt{scipy.integrate.odeint()} function (which
 internally uses~\texttt{lsoda} from~\texttt{ODEPACK}).
\end{itemize}
Concerning the numerical solution of the ODE, some manual tweaking of
integration parameters such as maximal step sizes is required when the
coupling constant~$g$ is large and~$k\sim m$. We note that, by the
very nature of this problem, the computation is readily parallelized:
The effort to numerically determine the right hand side integral is
expected to roughly grow like~$\mathcal O(N^2)$ with the number~$N$ of
support points, and computations for different support points are
independent. In comparison to the computational effort required to
compute the integrals, the communication overhead to distribute the
values of~$z(k;p)$ at different support points is fairly negligible,
hence using one of the readily available MPI-extensions to Python to
intelligently distribute the workload becomes an attractive
option. One should, however, take care that the core structure of the
integrand then is implemented in a compiled (C code) Python extension
before even thinking about parallelization.

\section{Determination of the renormalized mass}
\label{sec:DetRenMass}
\begin{figure*}
	\includegraphics[width=.45\textwidth]{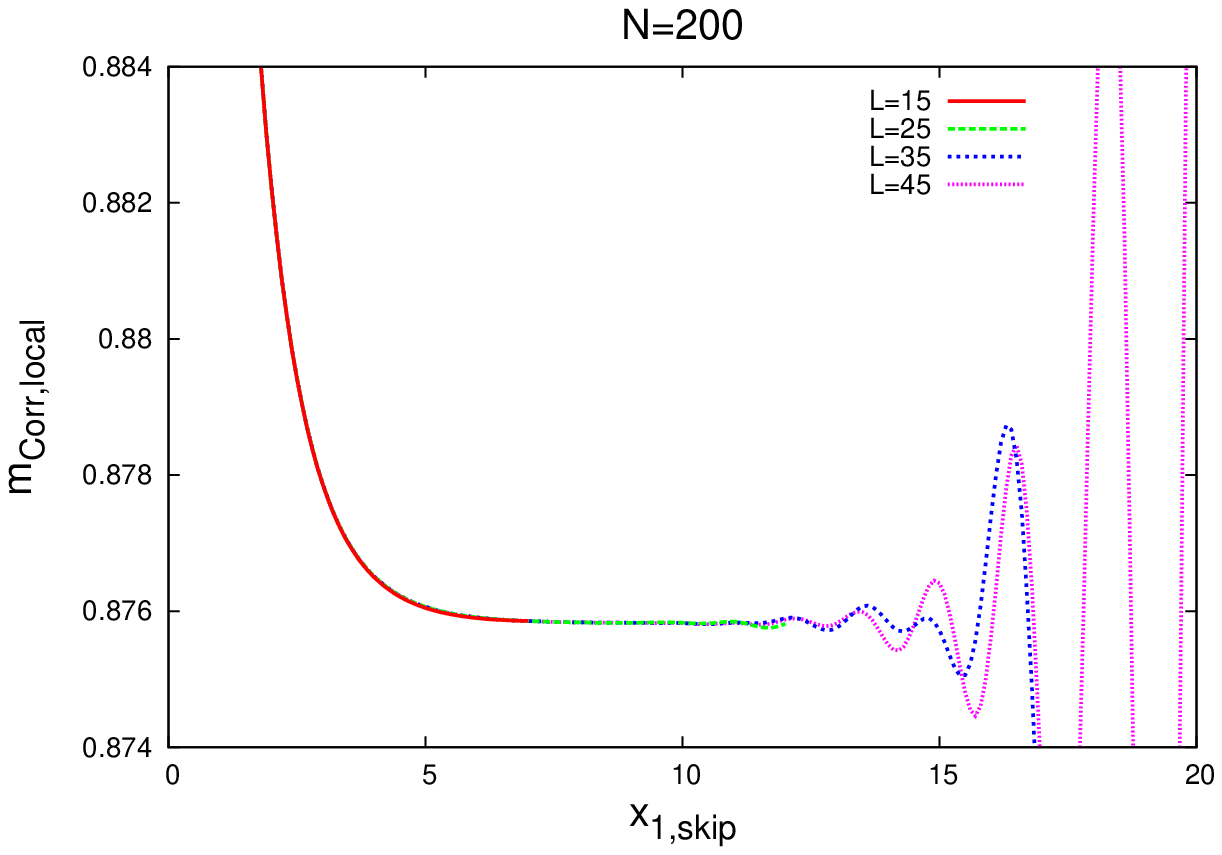}
	\includegraphics[width=.45\textwidth]{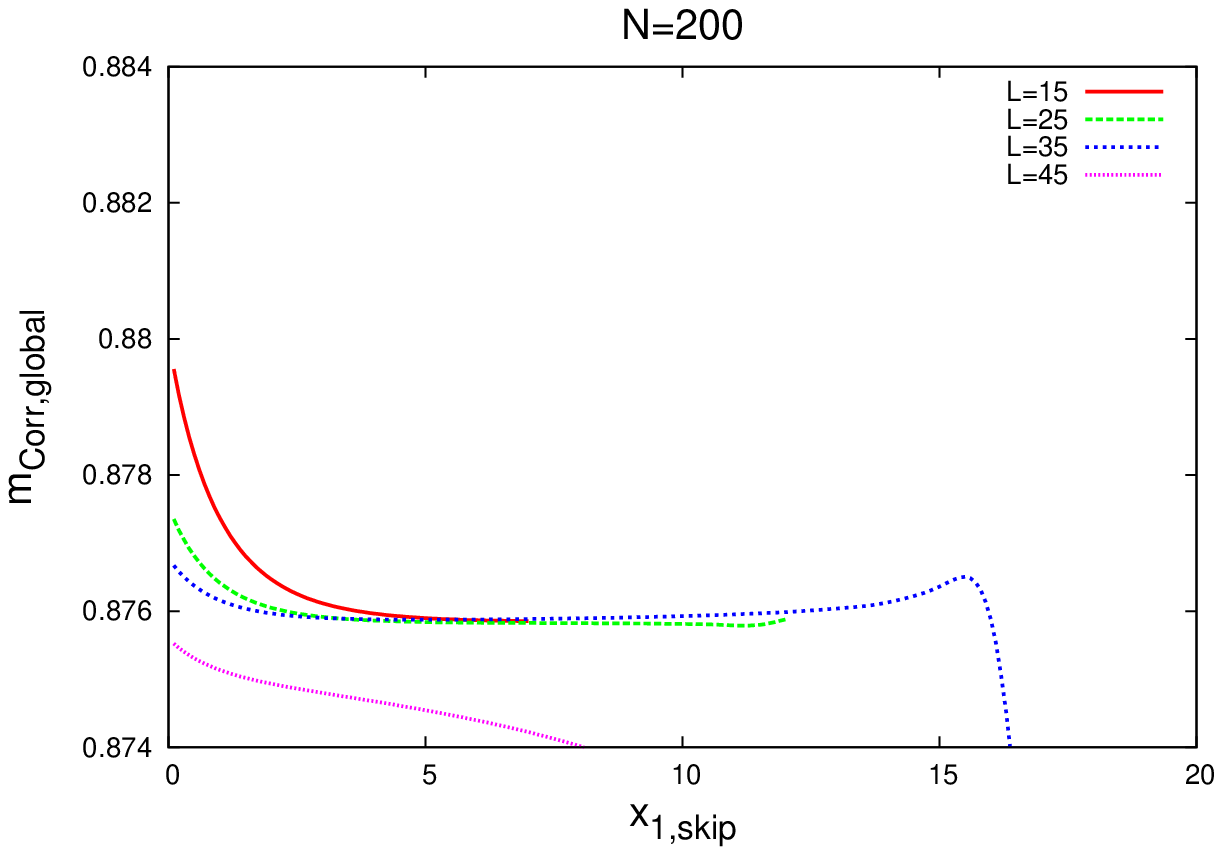}
	\includegraphics[width=.45\textwidth]{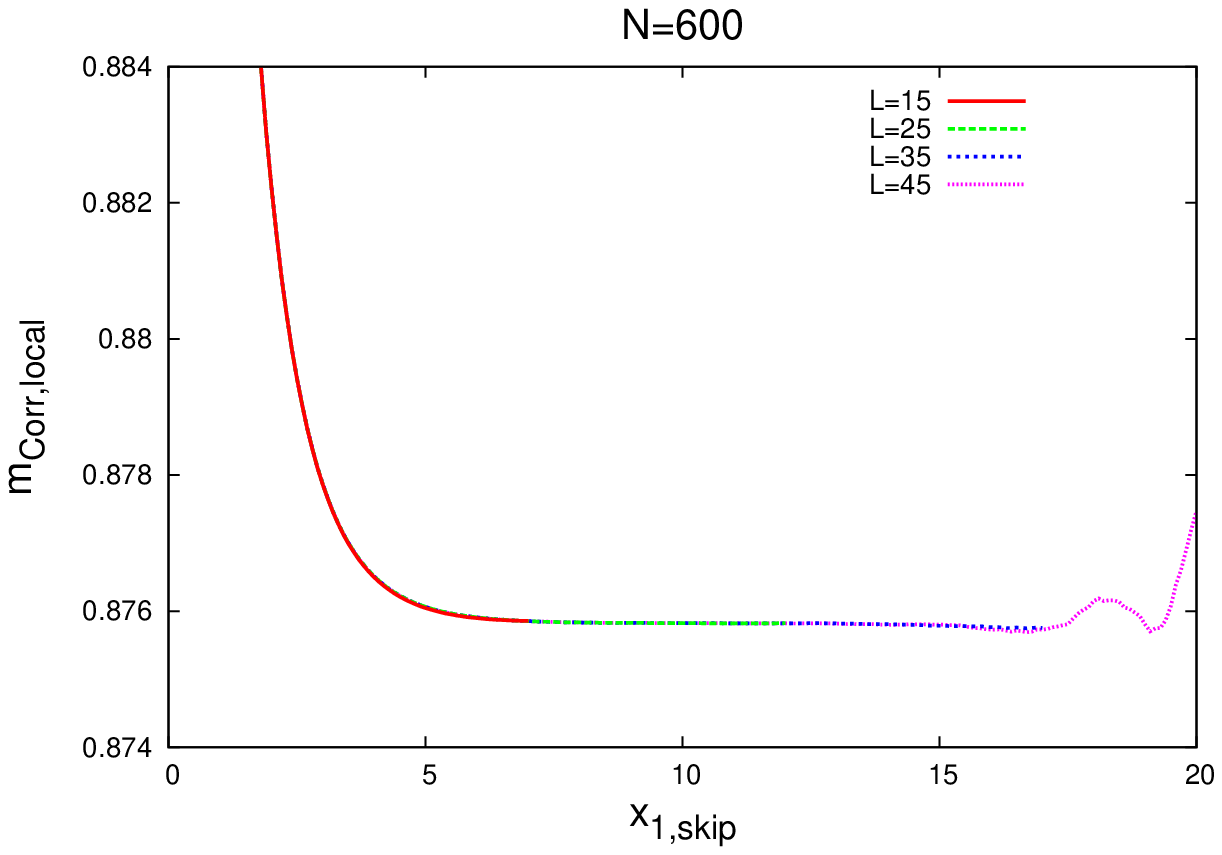}
	\includegraphics[width=.45\textwidth]{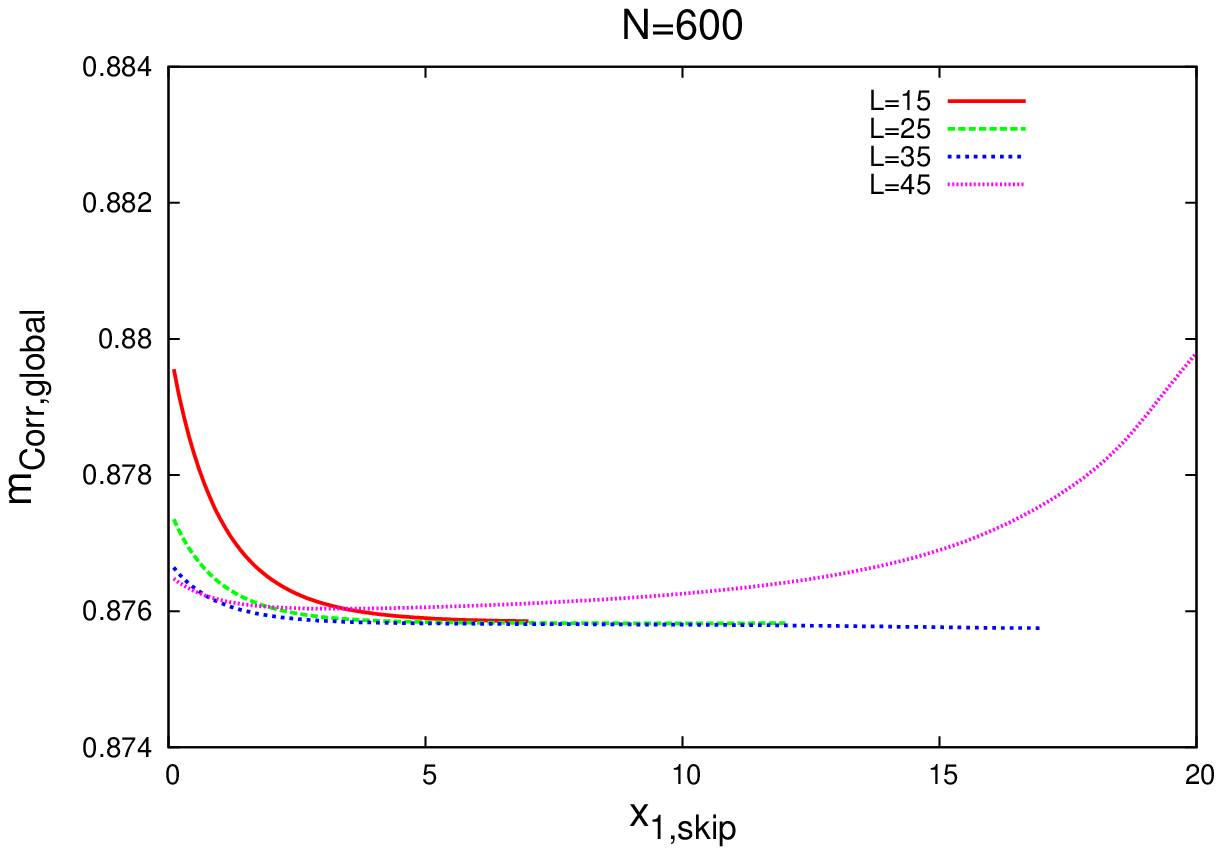}
	\caption{Left panel: $m_{\rm corr}^{\rm local}$ with $\lambda=0.6$ for
	discretizations $N=200$ and $N=600$ and different box sizes
	$L=15,25,35$ and $45$. Right panel: $m_{\rm corr}^{\rm global}$  with
	$\lambda=0.6$  for discretizations $N=200$ and $N=600$ and different box
	sizes $L=15,25,35$ and $45$.
	\label{fig:effMassLam06}}
\end{figure*}

The numerical calculations of $Z_k^2$ in the main text use a grid of $N=60$
points in the direction of $p^2$, distributed equidistantly on
a logarithmic scale. The result for 
$Z_{k\to0}(p)$ is interpolated with splines to calculate the propagator~$G_{\rm bos}^{\rm
NLO}(p)$. A discreet Fourier transformation of $G_{\rm bos}^{\rm
NLO}(p)$ yields the correlator  $C(x_1)$ on the interval $x_1\in[0,L]$ with
$n=10\,001$ intermediate points. In the main text we have used $L=15$. From its
 large distance behavior
 \begin{equation}
C_{a,m_{\rm cor}}(x_1)\propto\cosh(\mcor(x_1-L/2))\,  
\end{equation}
the correlator mass $m_{\rm corr}$ is determined by a least square fit. 
The fit range is constrained to the interval $[x_{1,skip},\ldots, L-x_{1,skip}]$ 
 where the contributions of excited states are negligible.  The value of
 $x_{1,skip}$ is determined such that $m_{\rm corr}(x_{1,skip})$ shows a plateau.  We can either
 fit on the whole range $[x_{1,skip},\ldots, L-x_{1,skip}]$ -- this quantity is
 called $m_{\rm corr}^{\rm global}$ -- or make the fit just inside a small
 interval of size 0.2 starting  from $x_{1,\rm skip}$ -- this
 quantity is called $m_{\rm corr}^{\rm local}$.
 
In the left panel of Fig.~\ref{fig:effMassLam06}  $m_{\rm corr}^{\rm
local}$ is shown for two different discretisations of $Z_k^2(p^2)$, $N=200$ in
the upper and $N=600$ in the lower panel. In the right panel the same is shown
for $m_{\rm corr}^{\rm global}$.
From these plots we can read off that
 for $x_{1,skip}$ not too large there is a clear plateau which is stable if
the box size is  increased. But
  for very large $x_{1,skip}$ the local mass oscillates. As this
 oscillation is reduced if the discretization is increased it is due to
 fluctuations in the spline interpolation of $Z_k^2$.
At small values of the correlator the numerical errors are more important for the masses. 
  As the fluctuations become visible for large box sizes, in these cases 
 the global mass fit is of no use because it averages over the local
 mass and is strongly influenced by the oscillations.
We will therefore take the plateau of $m_{\rm corr}^{\rm local}$ as the
value of the renormalized mass.

\section{Perturbation Theory}
\label{sec:PertTheory}
In perturbation theory the mass is determined from the one particle irreducible (proper) vertex $\Sigma(p)$.
The perturbative expansion of this vertex is
\newlength{\diagheight}
\setlength{\diagheight}{0.8cm}
\newlength{\shiftdiag}
\setlength{\shiftdiag}{-0.3cm}
\newcommand{\insertdiag}[1]{\raisebox{\shiftdiag}{\includegraphics[height=\diagheight]{#1}}}
\begin{align}
\insertdiag{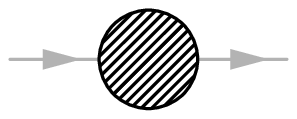}=&\insertdiag{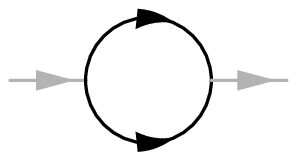}+\insertdiag{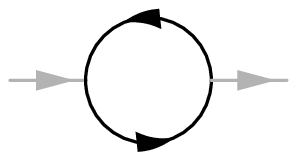}\\
\nonumber
&+\insertdiag{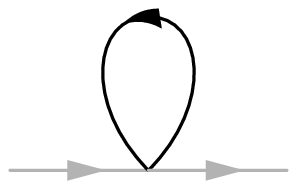}+\insertdiag{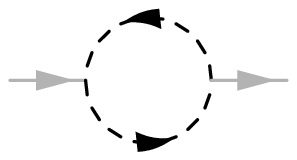}+O(g^4)
\end{align} 
\begin{align}
\nonumber
 \Sigma(p)=&2\gbar^2\int 
  \frac{d^2s}{4\pi^2}\frac{6\mbar^2-4((\vect
  p-\vect s)^2+\mbar^2)}{(s^2+\mbar^2)((\vect s-\vect p)^2+\mbar^2)}\\ 
\nonumber\displaybreak[0]
  &
   +2 g^2\int  
   \frac{d^2s}{4\pi^2}\frac{4(\vect
   s^2-\vect s\cdot \vect p-i\epsilon^{\mu\nu}\vect s_\mu(\vect
   s_\nu-\vect p_\nu))} {(s^2+\mbar^2)((\vect s-\vect p)^2+\mbar^2)}\\
\nonumber\displaybreak[0]  
   =&2
   g^2\int\frac{d^2s}{4\pi^2}\frac{2(\mbar^2-\vect
   p^2)}{(\vect s^2+\mbar^2)((\vect s-\vect p)^2+\mbar^2)}\\
\nonumber \displaybreak[0] 
  =&2\gbar^2\int_0^1dz
   \int\frac{d^2s}{4\pi^2}\frac{2(\mbar^2-p^2)}{(s^2+z(1-z)p^2+\mbar^2)^2}\\
\nonumber\displaybreak[0]
   =&2\gbar^2\int_0^1dz \frac{2(\mbar^2-p^2)}{4\pi(z(1-z)p^2+\mbar^2)^2}\\
   \displaybreak[0]
   =&
   \frac{4\gbar^2(\mbar^2-p^2)}{\pi p\sqrt{4\mbar^2+p^2}}\,
  \mathrm{artanh}
   \left({p({4\mbar^2+p^2})^{-1/2}}\right)
\end{align}



\begin{thebibliography}{10}

\bibitem{Catterall:2009it}
Simon Catterall, David~B. Kaplan, and Mithat Unsal.
\newblock {Exact lattice supersymmetry}.
\newblock {\em Phys.Rept.}, 484:71--130, 2009. 
\newblock arXiv:0903.4881 [hep-lat].

\bibitem{Giedt:2006pd}
Joel Giedt.
\newblock {Deconstruction and other approaches to supersymmetric lattice field
  theories}.
\newblock {\em Int. J. Mod. Phys.}, A21:3039--3094, 2006.
\newblock hep-lat/0602007.

\bibitem{Bergner:2007pu}
Georg Bergner, Tobias Kaestner, Sebastian Uhlmann, and Andreas Wipf.
\newblock {Low-dimensional supersymmetric lattice models}.
\newblock {\em Annals Phys.}, 323:946--988, 2008.
\newblock arXiv:0705.2212 [hep-lat].

\bibitem{Kastner:2008zc}
Tobias Kastner, Georg Bergner, Sebastian Uhlmann, Andreas Wipf, and Christian
  Wozar.
\newblock {Two-Dimensional Wess-Zumino Models at Intermediate Couplings}.
\newblock {\em Phys. Rev.}, D78:095001, 2008.
\newblock arXiv:0807.1905 [hep-lat].

\bibitem{Aoki:2000wm}
K.~Aoki.
\newblock {Introduction to the nonperturbative renormalization group and its
  recent applications}.
\newblock {\em Int. J. Mod. Phys.}, B14:1249--1326, 2000.

\bibitem{Berges:2000ew}
Jurgen Berges, Nikolaos Tetradis, and Christof Wetterich.
\newblock {Non-perturbative renormalization flow in quantum field theory and
  statistical physics}.
\newblock {\em Phys. Rept.}, 363:223--386, 2002.
\newblock hep-ph/0005122.

\bibitem{Litim:1998nf}
Daniel~F. Litim and Jan~M. Pawlowski.
\newblock {On gauge invariant Wilsonian flows}.
\newblock 1998.
\newblock hep-th/9901063.

\bibitem{Pawlowski:2005xe}
Jan~M. Pawlowski.
\newblock {Aspects of the functional renormalisation group}.
\newblock {\em Annals Phys.}, 322:2831--2915, 2007.
\newblock hep-th/0512261.

\bibitem{Gies:2006wv}
Holger Gies.
\newblock {Introduction to the functional RG and applications to gauge
  theories}.
\newblock 2006.
\newblock hep-ph/0611146.

\bibitem{Sonoda:2007av}
Hidenori Sonoda.
\newblock {The Exact Renormalization Group -- renormalization theory revisited
  --}.
\newblock 2007.
\newblock arXiv:0710.1662 [hep-th].

\bibitem{Rosten:2008ih}
Oliver~J. Rosten.
\newblock {On the Renormalization of Theories of a Scalar Chiral Superfield}.
\newblock \newblock {\em JHEP}, 1003:004, 2010.
\newblock arXiv:0808.2150 [hep-th].

\bibitem{Rosten:2010vm}
  O.~J.~Rosten,
  \newblock{Fundamentals of the Exact Renormalization Group}.
  \newblock arXiv:1003.1366 .
  
  \bibitem{Delamotte:2003dw}
  B.~Delamotte, D.~Mouhanna and M.~Tissier,
  \newblock{Nonperturbative renormalization group approach to frustrated
  magnets}.
  \newblock {\em Phys.\ Rev.}  B69:134413, 2004
  [arXiv:cond-mat/0309101].

\bibitem{Vian:1998kv}
F.~Vian.
\newblock {Supersymmetric gauge theories in the exact renormalization group
  approach}.
\newblock 1998.
\newblock hep-th/9811055.

\bibitem{Bonini:1998ec}
M.~Bonini and F.~Vian.
\newblock {Wilson renormalization group for supersymmetric gauge theories and
  gauge anomalies}.
\newblock {\em Nucl. Phys.}, B532:473--497, 1998.
\newblock hep-th/9802196.

\bibitem{Synatschke:2008pv}
Franziska Synatschke, Georg Bergner, Holger Gies, and Andreas Wipf.
\newblock {Flow Equation for Supersymmetric Quantum Mechanics}.
\newblock {\em JHEP}, 03:028, 2009.
\newblock arXiv:0809.4396 [hep-th].

\bibitem{Gies:2009az}
Holger Gies, Franziska Synatschke, and Andreas Wipf.
\newblock {Supersymmetry breaking as a quantum phase transition}.
\newblock {\em Phys. Rev.}, D80:101701(R), 2009.
\newblock arXiv:0906.5492 [hep-th].

\bibitem{Synatschke:2009nm}
Franziska Synatschke, Holger Gies, and Andreas Wipf.
\newblock {Phase Diagram and Fixed-Point Structure of two dimensional N=1
  Wess-Zumino Models}.
\newblock {\em Phys. Rev.}, D80:085007, 2009.
\newblock arXiv:0907.4229 [hep-th].

\bibitem{Synatschke:2010ub}
Franziska Synatschke, Jens Braun, and Andreas Wipf.
\newblock {N=1 Wess Zumino Model in d=3 at zero and finite temperature}.
\newblock 2010.
\newblock arXiv:1001.2399 [hep-th].

\bibitem{Falkenberg:1998bg}
Sven Falkenberg and Bodo Geyer.
\newblock {Effective average action in N = 1 super-Yang-Mills theory}.
\newblock {\em Phys. Rev.}, D58:085004, 1998.
\newblock hep-th/9802113.

\bibitem{Arnone:2004ey}
S.~Arnone and K.~Yoshida.
\newblock {Application of exact renormalization group techniques to the
  non-perturbative study of supersymmetric field theory}.
\newblock {\em Int. J. Mod. Phys.}, B18:469--478, 2004.

\bibitem{Arnone:2004ek}
Stefano Arnone, Francesco Guerrieri, and Kensuke Yoshida.
\newblock {N = 1* model and glueball superpotential from renormalization group
  improved perturbation theory}.
\newblock {\em JHEP}, 05:031, 2004.
\newblock hep-th/0402035.

\bibitem{Sonoda:2009df}
Hidenori Sonoda and Kayhan Ulker.
\newblock {An elementary proof of the non-renormalization theorem for the
  Wess-Zumino model}.
\newblock 2009.
\newblock arXiv:0909.2976 [hep-th].

\bibitem{Sonoda:2008dz}
Hidenori Sonoda and Kayhan Ulker.
\newblock {Construction of a Wilson action for the Wess-Zumino model}.
\newblock {\em Prog.Theor.Phys.}, 120:197-230, 2008.
\newblock arXiv:0804.1072 [hep-th].

\bibitem{Ellwanger:1995qf}
Ulrich Ellwanger, Manfred Hirsch, and Axel Weber.
\newblock {Flow equations for the relevant part of the pure Yang- Mills
  action}.
\newblock {\em Z. Phys.}, C69:687--698, 1996.
\newblock hep-th/9506019.

\bibitem{Pawlowski:2003hq}
Jan~M. Pawlowski, Daniel~F. Litim, Sergei Nedelko, and Lorenz von Smekal.
\newblock {Infrared behaviour and fixed points in Landau gauge QCD}.
\newblock {\em Phys. Rev. Lett.}, 93:152002, 2004.
\newblock hep-th/0312324.

\bibitem{Fischer:2004uk}
Christian~S. Fischer and Holger Gies.
\newblock {Renormalization flow of Yang-Mills propagators}.
\newblock {\em JHEP}, 10:048, 2004.
\newblock hep-ph/0408089.

\bibitem{Blaizot:2006vr}
Jean-Paul Blaizot, Ramon Mendez-Galain, and Nicolas Wschebor.
\newblock {Non perturbative renormalization group and momentum dependence of
  n-point functions. II}.
\newblock {\em Phys. Rev.}, E74:051117, 2006.
\newblock hep-th/0603163.

\bibitem{Blaizot:2005wd}
Jean-Paul Blaizot, Ramon Mendez-Galain, and Nicolas Wschebor.
\newblock {Non perturbative renormalisation group and momentum dependence of
  n-point functions. I}.
\newblock {\em Phys. Rev.}, E74:051116, 2006.
\newblock hep-th/0512317.

\bibitem{Blaizot:2005xy}
J.~P. Blaizot, Ramon Mendez~Galain, and Nicolas Wschebor.
\newblock {A new method to solve the non perturbative renormalization group
  equations}.
\newblock {\em Phys. Lett.}, B632:571--578, 2006.
\newblock hep-th/0503103.

\bibitem{Benitez:2009xg}
F.~Benitez et~al.
\newblock {Solutions of renormalization group flow equations with full momentum
  dependence}.
\newblock {\em Phys. Rev.}, E80:030103, 2009.
\newblock arXiv:0901.0128 [cond-mat.stat-mech].

\bibitem{Diehl:2007xz}
S.~Diehl, H.~C. Krahl, and M.~Scherer.
\newblock {Three-Body Scattering from Nonperturbative Flow Equations}.
\newblock {\em Phys. Rev.}, C78:034001, 2008.
\newblock arXiv:0712.2846 [cond-mat.stat-mech].

\bibitem{Wess:1973kz}
J.~Wess and B.~Zumino.
\newblock {A Lagrangian Model Invariant Under Supergauge Transformations}.
\newblock {\em Phys. Lett.}, B49:52, 1974.

\bibitem{West}
P.~West.
\newblock {\em Introduction to supersymmetry and supergravity}.
\newblock World Scientific, extended second edition, 1990.

\bibitem{Witten:1982df}
Edward Witten.
\newblock {Constraints on Supersymmetry Breaking}.
\newblock {\em Nucl. Phys.}, B202:253, 1982.

\bibitem{Witten:1993yc}
Edward Witten.
\newblock {Phases of N = 2 theories in two dimensions}.
\newblock {\em Nucl. Phys.}, B403:159--222, 1993.
\newblock hep-th/9301042.

\bibitem{MirrorSymmetry}
Kentaro Hori, Sheldon Katz, Albrecht Klemm, Rahul Pandharipande, Richard
  Thomas, Cumrun Vafa, Ravi Vakil, and Eric Zaslow.
\newblock {\em Mirror symmetry}.
\newblock Clay Mathematics Institute, Cambridge, MA, 2003.

\bibitem{Gates:1983nr}
S.~J. Gates, Marcus~T. Grisaru, M.~Rocek, and W.~Siegel.
\newblock {Superspace, or one thousand and one lessons in supersymmetry}.
\newblock {\em Front. Phys.}, 58:1--548, 1983.
\newblock hep-th/0108200.

\bibitem{Beccaria:1998vi}
Matteo Beccaria, Giuseppe Curci, and Erika D'Ambrosio.
\newblock {Simulation of supersymmetric models with a local Nicolai map}.
\newblock {\em Phys. Rev.}, D58:065009, 1998.
\newblock hep-lat/9804010.

\bibitem{Catterall:2003ae}
Simon Catterall and Sergey Karamov.
\newblock {A lattice study of the two-dimensional Wess Zumino model}.
\newblock {\em Phys. Rev.}, D68:014503, 2003.
\newblock hep-lat/0305002.

\bibitem{Giedt:2005ae}
Joel Giedt.
\newblock {R-symmetry in the Q-exact (2,2) 2d lattice Wess-Zumino model}.
\newblock {\em Nucl. Phys.}, B726:210--232, 2005.
\newblock hep-lat/0507016.

\bibitem{Wetterich:1992yh}
Christof Wetterich.
\newblock {Exact evolution equation for the effective potential}.
\newblock {\em Phys. Lett.}, B301:90--94, 1993.

\bibitem{Litim:2000ci}
Daniel~F. Litim.
\newblock {Optimisation of the exact renormalisation group}.
\newblock {\em Phys. Lett.}, B486:92--99, 2000.
\newblock hep-th/0005245.

\bibitem{Litim:2001up}
Daniel~F. Litim.
\newblock {Optimised renormalisation group flows}.
\newblock {\em Phys. Rev.}, D64:105007, 2001.
\newblock hep-th/0103195.

\bibitem{Litim:2001hk}
Daniel~F. Litim and Jan~M. Pawlowski.
\newblock {Predictive power of renormalisation group flows: A comparison}.
\newblock {\em Phys. Lett.}, B516:197--207, 2001.
\newblock hep-th/0107020.

\bibitem{Litim:2002cf}
Daniel~F. Litim.
\newblock {Critical exponents from optimised renormalisation group flows}.
\newblock {\em Nucl. Phys.}, B631:128--158, 2002.
\newblock hep-th/0203006.

\bibitem{Litim:2006ag}
Daniel~F. Litim and Jan~M. Pawlowski.
\newblock {Non-perturbative thermal flows and resummations}.
\newblock {\em JHEP}, 11:026, 2006.
\newblock hep-th/0609122.

  
\bibitem{Canet:2002gs}
  L.~Canet, B.~Delamotte, D.~Mouhanna and J.~Vidal,
  \newblock {Optimization of the derivative expansion in the nonperturbative
  renormalization group}.
  \newblock{Phys.\ Rev.}  D67:065004, 2003
  [arXiv:hep-th/0211055].
  



\bibitem{Gies:2001nw}
Holger Gies and Christof Wetterich.
\newblock {Renormalization flow of bound states}.
\newblock {\em Phys. Rev.}, D65:065001, 2002.
\newblock hep-th/0107221.

\bibitem{tbp}
Thomas Fischbacher and Franziska Synatschke.
\newblock work in progress.

\bibitem{SciPy}
\url{http://www.scipy.org/}.

\bibitem{Python}
\url{http://lucas.iquanta.info/linux/Python/docs/docs/official_2_5/ref.pdf}.

\bibitem{Bergner:2009vg}
Georg Bergner.
\newblock {Complete supersymmetry on the lattice and a No-Go theorem: A
  simulation with intact supersymmetries on the lattice}.
\newblock {\em JHEP}, 01:024, 2010.
\newblock arXiv:0909.4791.

\bibitem{Wozar}
G.~Bergner and C~Wozar.
\newblock in preperation.







\end{thebibliography}
\end{document}